\documentclass[10pt,aps,prl,reprint,showpacs,amsmath,amssymb]{revtex4-1}
\usepackage[pdfpagemode=None]{hyperref}
\usepackage{graphicx,bm}
\usepackage{color}
\usepackage{pdfpages}

\newcommand{\ket}[1]{\left| #1 \right>}
\newcommand{\Tc}{\ensuremath{T_\mathrm{c}}}

\begin{document}
\title{Fluctuational susceptibility of ultracold bosons  in the vicinity of condensation in the presence of an artificial magnetic field}

\author{A. J. Kingl} \affiliation{School of Physics and Astronomy, University
  of Birmingham, Edgbaston, Birmingham, B15 2TT, United Kingdom}
\author{D. M. Gangardt} \affiliation{School of Physics and Astronomy,
  University of Birmingham, Edgbaston, Birmingham, B15 2TT, United Kingdom}
\author{I. V. Lerner} \affiliation{School of Physics and Astronomy, University
  of Birmingham, Edgbaston, Birmingham, B15 2TT, United Kingdom}

\begin{abstract}
  We study the behavior of ultracold bosonic gases in the critical region above the Bose-Einstein condensation in the presence of an artificial magnetic field, $B_\mathrm{art}$. We show that the condensate fluctuations above the critical temperature \Tc\ cause the fluctuational susceptibility, $\chi_\mathrm{fl}$, of a uniform gas to have a stronger power-law divergence than in an analogous superconducting system. Measuring such a divergence opens new ways of exploring critical properties of   ultracold gases and  an opportunity of an accurate determination of \Tc. We describe a method of measuring $\chi_\mathrm{fl}$ which requires a constant gradient in $B_\mathrm{art}$ and suggest a way of creating   such a field in experiment.
\end{abstract}
\maketitle

Amongst intensive simulation of condensed-matter effects in cold atomic gases (see  \cite{Lewenstein:07,BlochZwerger:2008,Pfau:09} for reviews), considerable attention was focused both on similarities and on striking differences in properties of superconducting systems on the one hand and ultra-cold Bose systems on the other  (see \cite{Sheehy:07} for review).  Yet, the impact of  fluctuations of the condensate order parameter above a critical temperature $T_\mathrm{c}$   remains to be observed  in atomic gases.

In the vicinity of \Tc, i.e.\ for $|\tau|\ll1$ where $\tau\equiv{T}/T_\mathrm{c}-1$ is a reduced temperature, superconductivity can be described within the Ginzburg-Landau mean-field (MF)  theory \cite{Landau:50}. Its tremendous success  for conventional clean superconductors is based on irrelevancy of the fluctuations for all achievable temperatures due  the smallness of the Ginzburg number,  $\mathrm{Gi}\sim10^{-12}\!\div\!10^{-14}$. Here the Ginzburg number $\mathrm{Gi}$ defines the temperature interval, $|\tau|\lesssim\mathrm{Gi}$, where fluctuational effects dominate  \cite{Levanyuk:59,*Ginzburg:61}. However, $\mathrm{Gi}$ is much larger in dirty superconductors so that temperatures $\tau\sim\mathrm{Gi}$ become attainable.
In the temperature interval $\mathrm{Gi}\lesssim\tau\ll1$ the MF results still dominate but fluctuational corrections become observable and lead to a sharp power-law $\tau$-dependence of  conductivity \cite{Strongin:70} and magnetic response \cite{Yamaji:1971} \emph{above} \Tc. The observations made in  Refs.~\cite{Strongin:70,Yamaji:1971} were in excellent agreement with perturbative  predictions by Aslamazov and Larkin, Maki,  and Thompson \cite{AL:68a,*Maki:1968,*Thompson:70,LV:05}.

No similar observations exist for gases of cold bosons where analogs of the magnetic susceptibility and conductivity are not readily available for measurements.
 On the other hand, the Ginzburg number $\mathrm{Gi}\gtrsim1$ for a typical dilute cold bosonic gas: although  it is proportional to a small gas parameter, the  numerical coefficient is large, see Eq.~(\ref{Gi1}) below. This makes the order-parameter fluctuations above $\Tc$ strong and their effects potentially observable.

 In this Letter we analyze the fluctuational  contribution,
 $\chi_\mathrm{fl}$, to the  susceptibility of a cold bosonic cloud in an
 artificial magnetic field, $B_\mathrm{art}$, and suggest how to measure it.
 {Up to now experimental studies of properties of the BEC \emph{phase
     transition}  were mostly aimed at the divergent correlation length  \cite{Donner:2007, Dalibard:15}. Studying experimentally the critical susceptibility would allow one to measure another critical exponent thus building a more comprehensive picture of the phase transition.}

\begin{figure}[t]
\begin{center}
 \includegraphics[width=0.55\columnwidth]{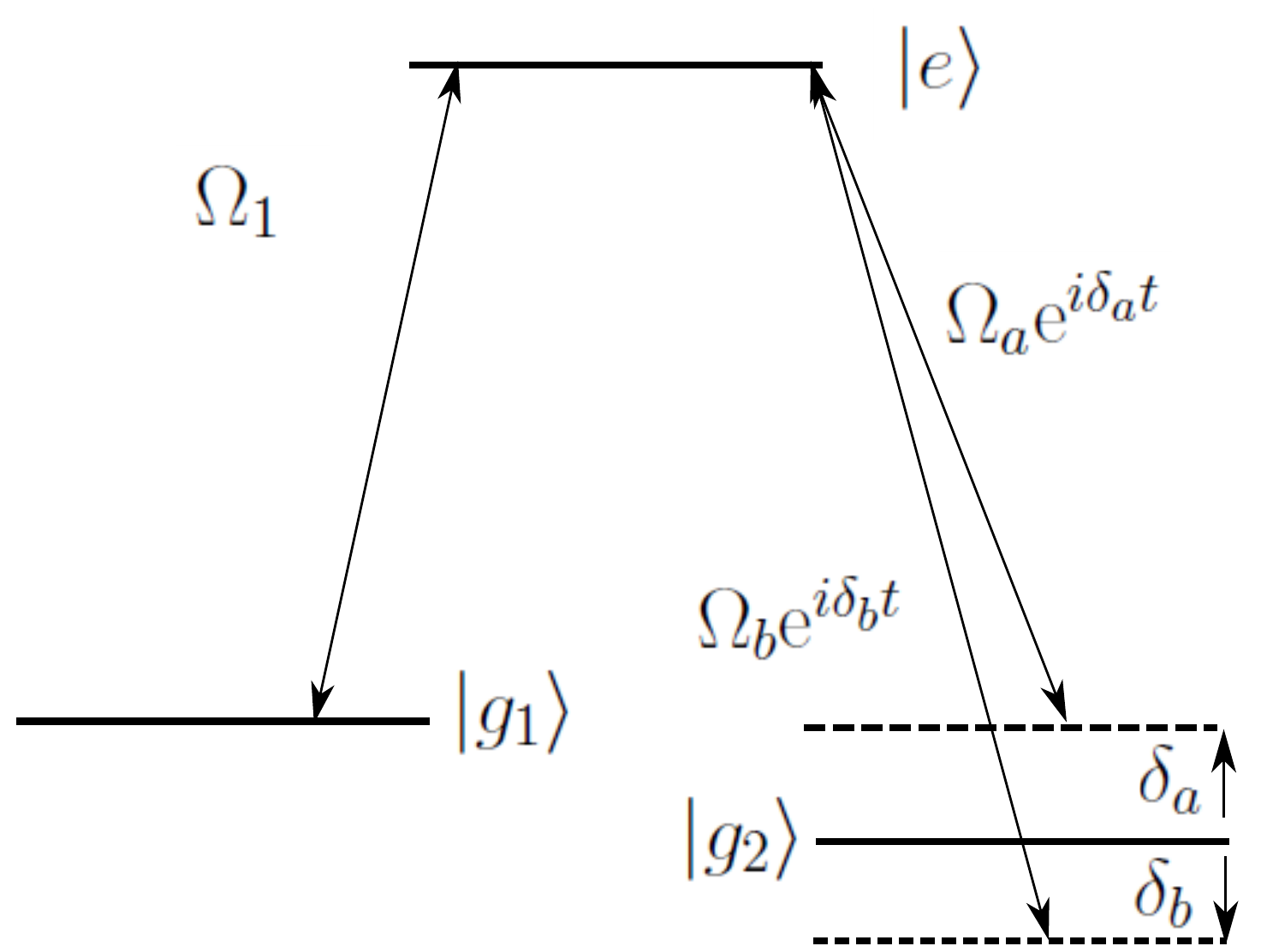}\\[16pt]
\includegraphics[width=0.55\columnwidth]{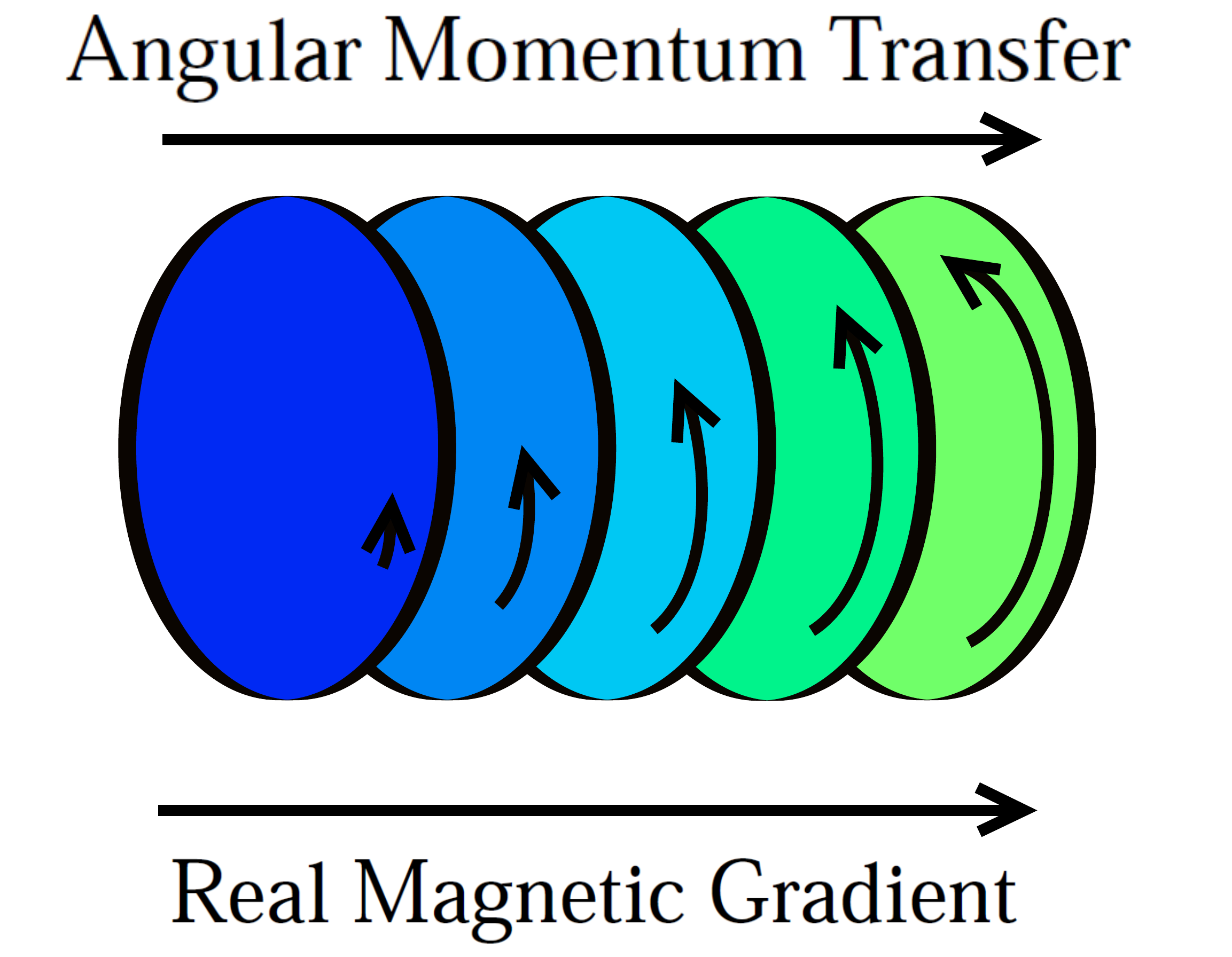}
\end{center}
\caption{The proposed setup of creating a
  nonuniform artificial  field,  $\bm{B}_{\mathrm{art}}$.
  Top panel -- the generalized $\Lambda$-scheme: torque is imprinted on the cloud by difference in angular momenta of three laser beams that couple sublevels of the ground state, $\ket{g_{1,2}}$ with exited state $|e\rangle$. Detunings of the beams coupling $\ket{g_2}$ and $\ket{e}$ change across the cloud due to a space-dependent Zeeman shift. Bottom panel -- varying detunings results in different parts of a bosonic cloud acquiring  different torques, which
 provides a gradient in the torque equivalent to $\nabla\bm{B}_\mathrm{art}$.}
\label{Fig:1}
\end{figure}

 We show that the dependence of $\chi_\mathrm{fl}$ on the reduced temperature
 $\tau$ is much sharper than in superconductivity  for the gas in a uniform
 trap, like that in Ref.~\cite{Hadzi,*hadzi2}. We argue that a realistic
 measuring scheme can be based on  using   field $B_\mathrm{art}$ with a
 constant gradient in space, and suggest a setup for creating such a field, see Fig.~\ref{Fig:1}. An implementation of such a scheme would expand the research in rotating condensates  \cite{Ketterle:01a,*Ketterle:01,*Gunn:02,*Cooper:01,*Ketterle:05,*Fetter:09}   and artificial  gauge fields in general  \cite{Osterloh:05,*Cooper:08,*Lin:09,*Kolovsky:11,Juzeliunas:04,*Dalibard:11,Brachmann:11}, which was mainly focused on increasing the flux densities to reach exotic states of matter, such as the quantum Hall regime  \cite{Sorensen:07,*Goldman:09,*Stanescu:10,*DemlerBloch:13}.

 An artificial magnetic field  $B_\mathrm{art}$ is created by imprinting an angular momentum and thus rotation on a cloud of neutral atoms.
  Neglecting interatomic interactions in the dilute   cloud above \Tc,  the corresponding  change in the free energy of the cloud of radius $R$ containing $N$ atoms is $F_0=-\frac{1}{4}mR^2N\omega_0^2$, which is equivalent to the free energy of rotation with frequency $\omega_0=\omega_B/2$ where $\omega_B\equiv{B}_\mathrm{art}/m$ is the analogue of the cyclotron frequency \cite{Legget}.  This looks like a rigid body rotation since the random thermal motion of atoms is averaged out.
 The susceptibility per particle   in natural units,
  \begin{align}\label{chi0}
     \chi_0&=-\frac{1}{N} \frac{\partial^2F_0}{\partial \omega_B^2}  = \frac{1}{8}mR^2,
  \end{align}
    is proportional to the average moment of inertia  per particle, which is reduced
below \Tc\ as the Bose-condensed part of the cloud does not contribute to it. The condensate fluctuations  above   \Tc\ result in  such a reduction, which we parameterize  as fluctuational corrections to the susceptibility: $\chi=\chi_0+\chi_\mathrm{fl}$.

A detection of such a  change in $\chi$ requires measuring the ratio of angular momentum to angular velocity  with a high precision. We suggest a scheme that removes the necessity of difficult (if at all possible) direct measurements by creating an artificial magnetic field  with a constant gradient in the direction of the field,  Fig.~\ref{Fig:1}.

The suggested scheme is based on modifying  the standard  $\Lambda$-scheme
\cite{Juzeliunas:04,*Dalibard:11,Brachmann:11}. There the electronic ground state of an atom  is Zeeman-split into two close
 sublevels,
  $\ket{g_{1,2}}$  with energies $\varepsilon_1\!\approx\!\varepsilon_2$  coupled to a single excited state, $\ket{e}$, of energy $\varepsilon_e$   by two
laser beams, with the second being   detuned  by frequency $\delta$ from the resonance $(\varepsilon_{e}-\varepsilon_2)/\hbar$. Angular momenta  $\ell_{1,2}$ are imprinted  (e.g., with holographic masks as in Ref.~\cite{Brachmann:11})  on each beam
so that their cross-sections have a Gaussian--Laguerre form with the  Rabi frequencies  adiabatically dependent on the atomic position, $\Omega_{1,2}({\bm{r}})=|\Omega_{1,2}|(\rho/\rho_0)^{\ell_{1,2}} e^{-\rho^2/2\rho_0^2}e^{i\ell_{1,2}\phi}$. We put  $|\Omega_1|\!\approx\!|\Omega_2|\!\equiv\Omega$, and $\ell_2\!\equiv\!\ell>0$ while $\ell_1\!=\!0$.
 Then  the coefficients of the internal atomic wave function, $\ket{\psi}=b_1(t)e^{-i\varepsilon_1t/\hbar} \ket{g_1}+b_2(t)e^{-i\varepsilon_2t/\hbar}\ket{g_2}+b_e(t)e^{-i\varepsilon_e t/\hbar}\ket{e}$, in the rotating wave approximation  obey the  equations
\begin{align} \label{Rabi}
    i\dot b_{1,2}({t})&=\Omega_{1,2}b_e({t})\;,
   &  i\dot b_e({t}) &= \Omega_1^*b_1({t})+\Omega_2^* b_2({t})\,.
\end{align}
In a steady state regime,   each atom finds itself in the dark state \cite{dark}, $\ket{d({\bm{r}})}=(\Omega_1({\bm{r}})\ket{g_1}-\Omega_2({\bm{r}})\ket{g_2})/(\Omega\sqrt{2})$, which is not directly coupled to the laser fields.
Two other internal states orthogonal to $\ket{d({\bm{r}})}$ are separated by a large gap and become redundant. A one-component wavefunction describing motion in the laser fields obeys a one-particle Schr\"{o}dinger equation {\cite{Juzeliunas:04,*Dalibard:11}} in a vector potential
$\bm{A}_\mathrm{art}=i\hbar \left<d|\bm\nabla d\right>$, corresponding to the artificial magnetic field in $z$-direction  \cite{Note5}
\begin{gather}\label{eq:qmagnet}
B_\mathrm{art} =|\bm\nabla \times \bm{A}_\mathrm{art} | =\frac{2\hbar \ell^2}{\rho_0^2} \frac{(\rho/\rho_0)^{2(\ell  -1)}}{ [{1+(\rho/\rho_0)^{2\ell}}]^2}\, f({\delta /\Omega})\,.
\end{gather}
Function $f({\delta/\Omega})$ describes the field sensitivity to detuning; it equals $[1+\delta^2/2\Omega^2]^{-1}$ for $|\delta|\ll\Omega$ and $2\Omega^2/\delta^2$ for $|\delta|\gg\Omega$ while its exact form is not relevant.

To create a gradient in $B_\mathrm{art}$, we suggest to modify the standard scheme by coupling $\ket{g_2}$ and $\ket{e}$ with two, instead of one, laser beams
 carrying different angular momenta, $\ell_{a,b}$    and  detuned   {by $\delta_{a,b}$} from the
 resonance, see top panel in  Fig.~\ref{Fig:1}. The gradient   arises from    linearly varying  the Zeeman  split  (with a weak \emph{real} magnetic field gradient  in the $z$-direction)  between $\ket{g_{1,2}}$ and thus the ratio $\delta_a/\delta_b$, resulting in a different angular momenta transfer to different cross-sections along the beams.

A rigorous  description of the modified scheme amounts to replacing   $\Omega_2$  in Eq.~(\ref{Rabi}) by $\Omega_2({t})=\Omega_ae^{i\delta_at}+\Omega_be^{i\delta_bt}$, where $\Omega_{a,b}(\bm{r})$ have the
Gaussian--Laguerre form  characterized by $\ell_{a,b}$.
\begin{figure}[b]
\includegraphics[width=0.9\columnwidth]{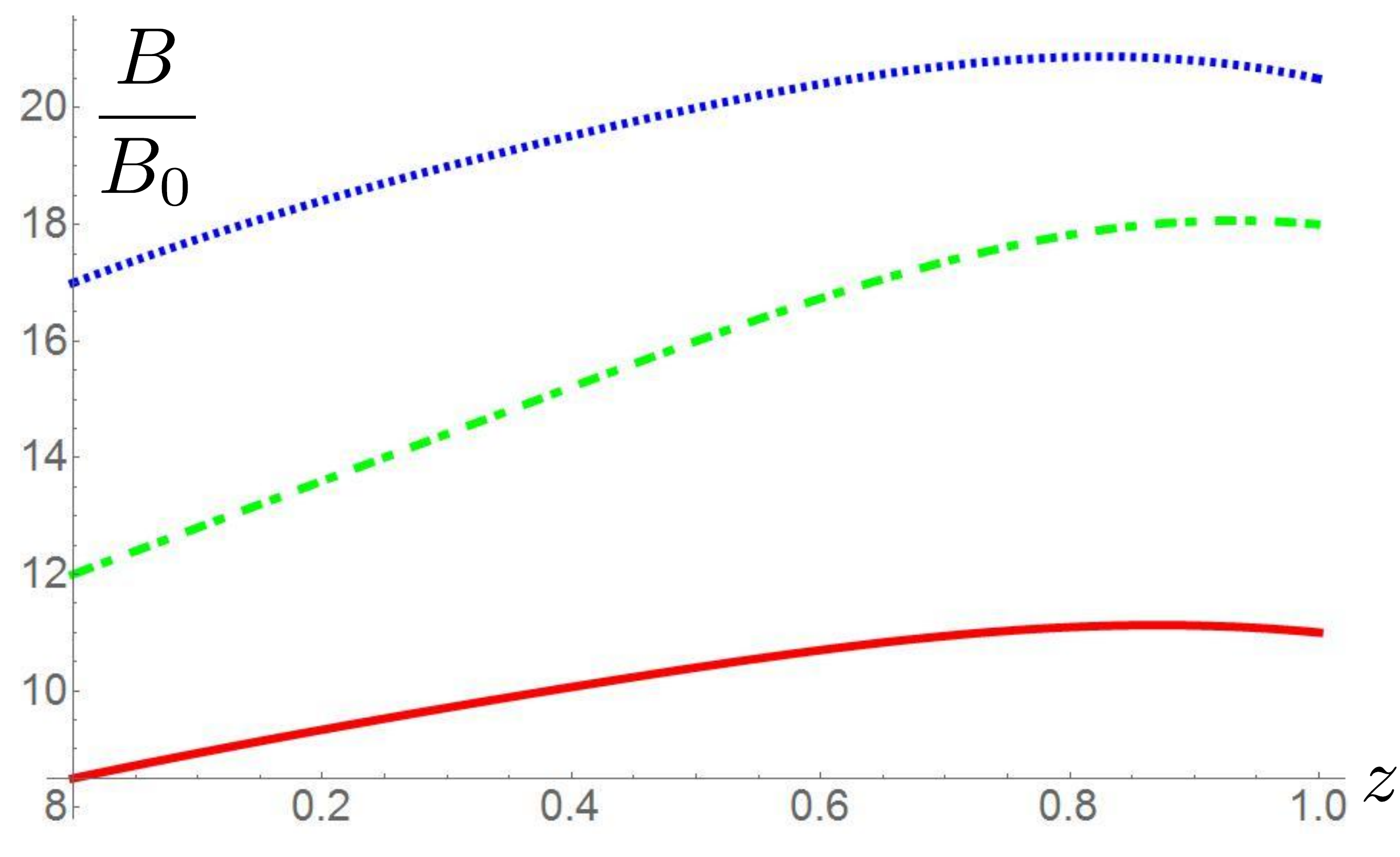} \caption{(color online) The artificial magnetic field $B_\mathrm{art}$ along
  the light-propagation axis $z$. The field is measured in units of
  $B_0={\hbar}/{\rho_0^2}$; the detuning  is chosen so that
   $\delta_a=0$ at $z=0$,  $\delta_b=0$ at
  $z=1$  and  $\delta_a-\delta_b=\Omega$ ($z$ is in arbitrary units).  Each line represents a different
  combination of angular momenta:  solid (red) is for $\ell_a=2$ and
  $\ell_b=3$;   dash-dotted (green) is for $\ell_a=2$ and $\ell_b=4$;
  dashed (blue) is for $\ell_a=3$ and $\ell_b=4$.}
\label{Fig:4}
\end{figure}
\begin{figure}[t]{\includegraphics[width=0.9\columnwidth]{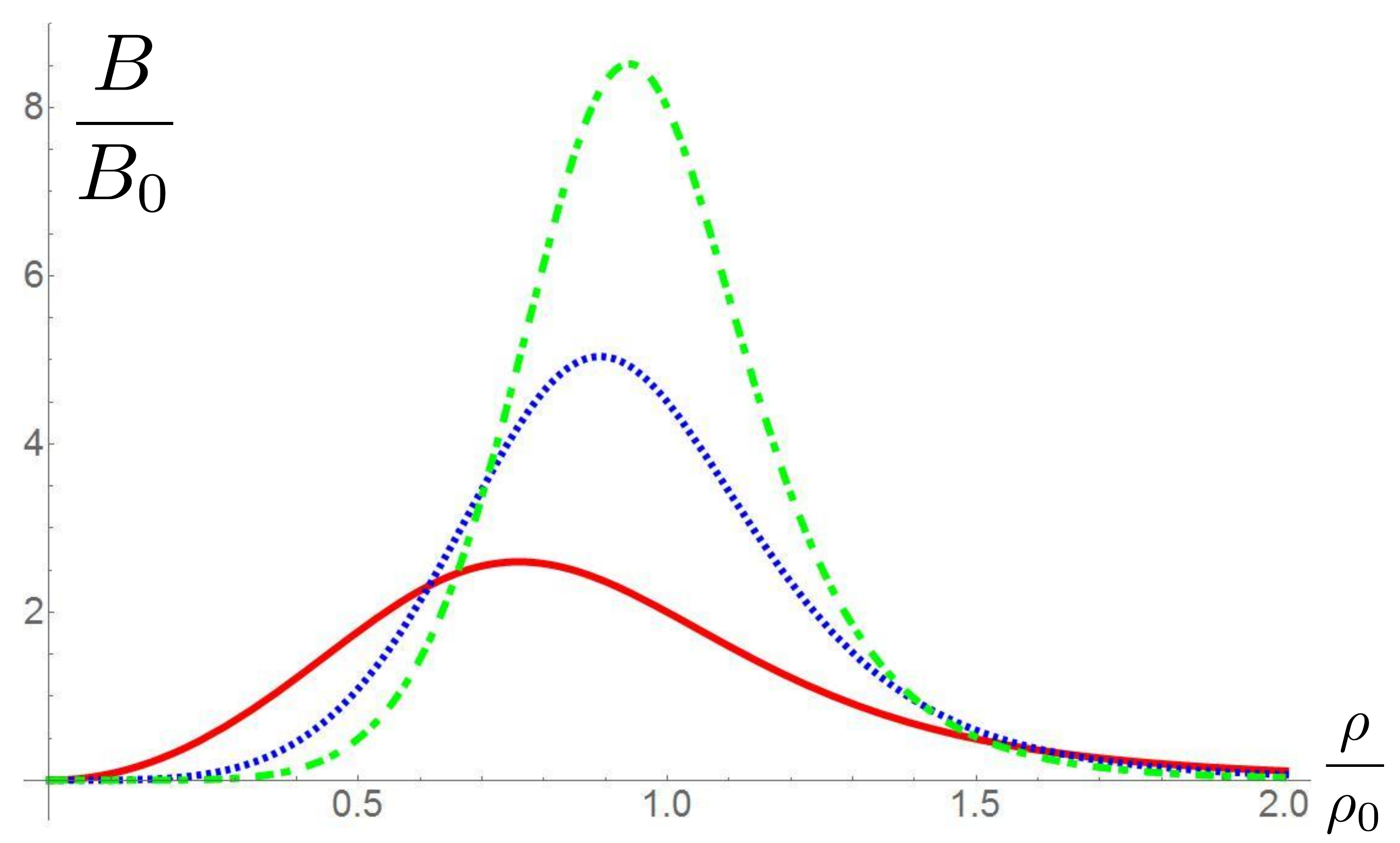}}
\label{Fig:5}\caption{(color online)  The radial
  dependence, $B_\mathrm{art}({\rho})$ of Eq.~\eqref{eq:qmagnet}, for $\Omega_b=0$ and
  zero detuning at different values of $\ell_a$: solid (red): $\ell_a=2$; dashed (blue): $\ell_a=3$;  dash-dotted (green): $\ell_a =4$. }
\end{figure}
One cannot eliminate time dependence from thus modified Eqs.~(\ref{Rabi}). However,  we can demonstrate  without exactly solving Eqs.~(\ref{Rabi})  that   an almost constant gradient of  $B_\mathrm{art}$ emerges with a proper choice of the parameters, as illustrated in Fig.~\ref{Fig:4}.  Such a gradient results from  setting the detunings in such a way that  (i) $\delta_a=0$ and $|\delta_b|\sim\Omega$ in the middle of the cloud  and (ii)  similar transverse profiles for $\Omega_a$ and $\Omega_b$   are selected by choosing $\ell_a>\ell_b>1$ (see Fig.~\ref{Fig:5}). In the Supplemental Material, we illustrate that the values chosen for this figure are optimal for making this constant gradient robust with respect to small changes in the applied real magnetic field and laser frequencies.

Now we  describe a response of the bosonic cloud with interatomic interactions to $B_{\mathrm{art}}$ created in the above scheme.  As the interactions  are typically too weak to excite atoms from the dark state $\ket d$ \cite{Note6},  an effective field theory can be formulated in terms of the one-component complex field $\Psi $ corresponding to $\ket d$. The action in  the vicinity of \Tc\ is    obtained from the full  imaginary time  action by retaining only
 the zeroth Matsubara-frequency component, $\Psi=\Psi({\bm{r}, \omega_n=0}) $, as higher frequencies do not contribute to singularities in $\tau$ \cite{Baym:99}. Thus results in \cite{Baym:99,Arnold:01,*ProkSvis:01}  the  universal classical action
\begin{align}\label{GL}
S [{\Psi}] =\frac{1}{T} \int\!\! d\bm{r} \biggl({\frac{\hbar^2|\bm\nabla \Psi|^2}{2m}-\delta\mu|\Psi|^2+\frac{g}{2}|\Psi|^4}\biggr)\,.
\end{align}
   Here  $\delta\mu =\mu-2gn $,  $\mu$ is  the chemical potential,  $n$ is  the particle density at the center of the trap,    $g=4\pi \hbar^2 a/m$, and $a$ is the scattering length. We assume weak interactions,  $an^{1/3}\ll1$. The coupling constant $g$ is practically unaffected by the $\Lambda$--scheme, which results only in changing the action by the minimal substitution, $\bm\nabla\to\bm\nabla-i \bm{A}_\mathrm{art}$, to include $B_\mathrm{art}$ \cite{Note5}.

Essentially,  $S[{\Psi}]$     in Eq.~(\ref{GL}) is  the standard
     Ginzburg-Landau  functional \cite{Note2}.
However, for weakly interacting bosons in a homogeneous  trap  a $\tau$-dependence of
fluctuations is sharper than in the  superconducting second order phase
transition.
Below \Tc\ the chemical potential is fully governed by the interaction:
$\delta\mu=-(3/p)gn \tau$,  where   $p=2,1$ for a  3D gas in a
homogeneous/harmonic  trap. On the contrary,  $\delta\mu$  above \Tc\ \cite{Note1} is essentially the same as for the ideal gas:
\begin{align}\label{mu}\delta\mu=\left\{%
      \begin{array}{ll}-c_1\Tc\tau,& \hbox{trapped gas;} \\-c_2 \Tc\tau ^2, & \hbox{uniform gas.}
      \end{array}\right.
\end{align}
where $c_1\approx 2.2$ and $c_2\approx 1.2$ \cite{Note3}.
The interaction correction to this  is parametrically small for the
 harmonically trapped gas where $gn\tau/|\delta\mu|\sim an^{1/3}\ll1$, and
 numerically small in the region of interest for the uniform gas where
 $gn\tau/|\delta\mu|<gn/(\Tc\mathrm{Gi})\sim0.1$, see  Eq.~(\ref{Gi1}) below.

 To show this, we rescale $\bm{r}\rightarrow\lambda\bm{r}$ and
 $\Psi\rightarrow{b}\Psi$ to make the coefficients attached to all the three
 terms in the GL functional (\ref{GL}) equal $1$.  The
 fluctuational weight ${e}^{-S[{\Psi}]/\Tc}$ becomes
 ${e}^{-\varkappa\widetilde{S}}$ where $\widetilde{S}$ is dimensionless and
 $\varkappa=(2\sqrt{|\delta\mu|}/{g\Tc})(\hbar^2/2m)^{3/2}\!$.  The
 Ginzburg criterion for suppression of the fluctuations \cite{Levanyuk:59}
 becomes $\varkappa\gtrsim1$. Substituting $\delta\mu$ and \Tc, we find
 $\varkappa\equiv({\tau/\mathrm{Gi}})^{p/2}$. Thus for the most interesting
 case of the 3D gas in a homogeneous trap ($p=2$) the Ginzburg criterion can
 be written as
 \begin{align}\label{Gi1}
1\gg|\tau |\gtrsim\mathrm{Gi}\approx
                       30 a n^{1/3}.
\end{align}
This coincides  up to a numerical prefactor with the condition earlier formulated in the cold-atom context \cite{Giorgini:1996,Donner:2007}.

The gas parameter $an^{1/3}$ is not small enough in typical dilute gases  to overcome the prefactor in Eq.~(\ref{Gi1}). For example, for a typical density of  trapped  Rb  atoms,   $10^{12}\div10^{13}cm^{-3}$ \cite{BlochZwerger:2008},  $an^{1/3}$  is   a few hundredths and $\mathrm{Gi}$ is just under $1$. However,  fine tuning the scattering length near a Feshbach resonance would allow one to reduce $\mathrm{Gi}$  by at least an order of magnitude \cite{Note7}, thus making the window ({\ref{Gi1}})  available for observations.

We consider cold atoms trapped in an optical lattice forming a stack of $N_l$
layers, where fluctuational effects are
stronger than in the bulk. We assume that the laser beams and thus
$\bm{B}_\mathrm{art}$ are normal to the layers. Then the minimal substitution
affects only in-layer components of the gradient term in Eq.~(\ref{GL}),
$\hbar\bm\nabla_\|\to\hbar\bm\nabla_\|-i\bm{A}_\mathrm{art}$, while the normal
component is replaced by $J|\Psi_j-\Psi_{j+1}|^2$, where $j$ enumerates layers
separated by a distance $d$ and $J$ is a weak inter-layer coupling.  After
integrating ${e}^{-S}$ over the fields $\Psi$, one finds \cite{LV:05,Note5}
the fluctuation contribution to free energy at $\tau\!\ll\!1$ as follows
\begin{gather*}%\label{Ffl}
F_\mathrm{fl}= \frac{\Phi \Tc}{\Phi_0}\sum_{n, k_z}\ln   \frac{\pi \Tc}{{\delta\mu( T)} + {\hbar \omega_B} (n+\tfrac{1}{2} )+ {4J} {\sin^2(\frac{k_z d}2)}}.
\end{gather*}
Here $\Phi$ is the total flux of $\bm{B}_\mathrm{art}$ through the layer,
$\Phi_0=2\pi\hbar$ plays the role of the elementary flux,  $n=0,1,2,\ldots$ labels the effective Landau levels in the field $B_\mathrm{art}$ and $k_z$ is a quasi-momentum in the normal direction.
The corresponding  susceptibility per particle, $\chi_\mathrm{fl}=-\frac{1}{N}{\partial^2F}/{\partial\omega_B^2}$,  is found  in the weak-field limit \cite{Note4} similarly to that for  superconductivity \cite{LV:05,Note5,LVV:08,*GVV:12}:
 \begin{gather}\label{chi}
    {\chi_\mathrm{fl}}/{\chi_0}  =  - ({2c_p}/{3 N_\|})
       {[{\tau^p ({\tau^p\!+\!\eta_p})}]}^{-\frac{1}{2}}  \,.
 \end{gather}
 For the gas in a homogeneous trap ($p=2$) the $\tau$ dependence  much sharper than in superconductivity results from the $\tau^2$-dependence of $\delta\mu$. Here   $N_\|\equiv{N}_\|({\rho_0})=\pi nd \rho_0^2$ is the number of particles in a magnetized part of a single layer and  $\eta_{1,2}={4J}/(c_{1,2}\Tc)$ is the anisotropy parameter.
 Since the inter-layer coupling $J$ is independent of other parameters, both  the regimes $\eta<\mathrm{Gi}$ or  $\eta>\mathrm{Gi}$ are possible.  In the latter case  a crossover between $2D$ and $3D$ behavior ($\tau >\eta$ or $\tau<\eta$) lies in the region of the MF applicability, Eq.~(\ref{Gi1}).

 The fluctuations susceptibility in Eq.~(\ref{chi})   is negative. Since the classical susceptibility $\chi _0$ is proportional to the moment of inertia,   a reduction of the overall susceptibility in the critical region  above \Tc\ is a fluctuational precursor of  the nonclassical rotational inertia below \Tc\ (the Hess--Fairbanks effect, \cite{Legget}). Similarly to superconductivity, where the reduced  magnetic susceptibility is a fluctuational precursor of the Meissner effect, such a reduction reflects  the divergence  of the size  {of} the fluctuational superfluid    droplets at \Tc.  A similar reduction of $\chi $ due to onset of superfluidity has recently been proposed \cite{cooper-hadzi} for measuring the superfluid fraction  below \Tc.

 Although the prefactor in Eq.~(\ref{chi})  is small, a very sharp $\tau$ dependence, especially in the case of the uniform gas, $\chi_\mathrm{fl} \propto \tau^{-2} $ at $\tau\gg\eta$  (which is much sharper than in superconductivity), makes the fluctuational effects observable. Such a sharp $\tau$ dependence should be even more pronounced outside region (\ref{Gi1}), i.e.\ for  $\tau<\mathrm{Gi}$, where analytical expression ({\ref{chi}}) is no longer valid but one still expects a critical behavior of $\chi$.  In this case the  appropriate critical exponent can   be in principle calculated numerically, as in the case of the critical correlation length \cite{ProkSvis:04,*Burovski:06,*Bezett:09,*Campostrini:09}
 where it turned out to be in an excellent agreement with the experiment \cite{Donner:2007}.

The next step is expressing $\chi_\mathrm{fl}$ via observable quantities. To increase the weight of $\chi_{\mathrm{fl}}$ in Eq.~(\ref{chi}), $N_\|$ and thus the laser beams aperture $\rho_0$ should be relatively small. On the other hand,  to ensure the linear response regime w.r.t.\ $B_\mathrm{art}$ it should be large enough, $\rho_0 n^{1/3}\gg1$, i.e.\ $\hbar\omega_B\sim\hbar^2/m\rho_0^2\ll\Tc$, or equivalently $\omega_B\rho_0\ll({\Tc/m})^{1/2}\sim v_T$. Taking $n\approx2.3\times10^{13}\mathrm{cm}^{-3}$ as in  measurements of the critical correlation length of a Rb cloud \cite{Donner:2007}, one can choose  $\rho_0\approx10n^{-1/3}\approx3\mu $m, (corresponding to $\omega_B\approx30$Hz) to  satisfy both the conditions.

In a steady state the central part of each layer rotates (after averaging out atomic thermal motion) with its own
angular velocity $\omega_0=\omega_B/2$, which  linearly changes from layer to layer due to the gradient of $B_\mathrm{art}$.
 After switching $B_\mathrm{art}$ off, it is necessary to allow some time for the equilibration within each layer, i.e.\ for redistributing the angular momentum from the central, ``magnetized" part across the layer by thermal collisions between particles.   The collision time, $\tau_\mathrm{col}\approx(a^2nv_T)^{-1}$, can be expressed via \Tc\ and $\mathrm{Gi}$, Eq.~(\ref{Gi1}), so that in uniform gas $\tau_\mathrm{col}\approx({\hbar/\Tc})({30/\mathrm{Gi}})^2$. This is about 30s if  the scattering length  $a$ is tuned so that  $\mathrm{Gi}\approx0.1$. So one should use the Feshbach resonance again to temporarily increase $a$ in order to facilitate the angular momentum redistribution.

 Expressing the angular momentum of the central part of the layer in terms of $\omega_B$ using   Eqs.~(\ref{chi0}) and (\ref{chi}), we find the angular velocity of each layer proportional to the  field $B_\mathrm{art}$ in this layer as follows
 \begin{gather*}%\label{omega}
    \omega({\tau})=\frac{N_\|({\rho_0})\chi ({\rho_0})}{N_\|({R})\chi ({R})}\frac{\omega_B}{2}=\frac{\rho_0^4}{R^4}\left[1+\frac{\chi_\mathrm{fl}({\rho_0,\tau})}{\chi_0({\rho_0})} \right]\frac{\omega_B}{2},
 \end{gather*}
 where the fluctuational corrections are included only in $\chi({\rho_0})$ as their relative contribution is much smaller in $\chi({R})$.  Noticing that $\chi_\mathrm{fl}$ is negligible at $\tau\sim1$ and $\omega_B$ is $T$-independent, we find that in each layer
 \begin{gather}\label{omega1}
    {\omega({\tau})}/{\omega({1})}=1+ {\chi_\mathrm{fl}({\rho_0,\tau})}/{ \chi_0({\rho_0})}
 \end{gather}
To extract  $\chi_\mathrm{fl}$ one needs to measure $\omega$ with a high precision, while keeping $\omega_B$ under control.  Having a constant gradient in $B_{\mathrm{art}}$, which makes each layer to rotate at different frequencies,  achieves precisely that.

To detect and measure such a differential rotation one can apply a short laser
pulse to make each layer elongated, as in the superfluid case
\cite{Dalibard:00}.  This elongation is preserved if the dephasing time due to
atomic collisions is much longer than the rotation period. Restoring the
scattering length to the value corresponding to $\mathrm{Gi}\approx0.1$
gives $\tau_\mathrm{col}\approx30$s that would preserve the shape for
hundreds of rotations with $\omega_B\approx30$Hz as above. Then after rotating
for time $t_0\sim\pi/\Delta\omega_0$, where $\Delta\omega_0$ is a difference
in angular velocities of the two outer layers, the orientations of all layers will
be uniformly distributed over all angles and the projection of the entire
cloud along the symmetry axis will change  from the elongated to the round
one.  The $T$-dependence $t_0({\tau})$ can be
found by repeated measurements of optical  density along the $z$ axis at
different temperatures in the vicinity of $\Tc$.  Then, inverting
Eq.~(\ref{omega1}), one expresses the fluctuations susceptibility in terms of
directly measurable dephasing times as
${\chi_\mathrm{fl}}/{\chi_0}=1-{t_0({\tau})}/{t_0({1})}$, which should reveal
the critical temperature dependence (\ref{chi}). An additional experimental
control may be achieved by measuring the revival time $N_lt_0$ when all the
elongated layers are aligned again.

 In conclusion, we have shown that an impact of the order parameter
 fluctuations on properties of ultracold bosonic systems at the onset of the
 BEC could be experimentally accessible.  In contrast to superconductivity,
 the fluctuational susceptibility is fully described \cite{Note2} by the GL
 functional (\ref{GL}) if the Ginzburg number (\ref{Gi1}) is small enough.
 Thus measuring $\chi_{\mathrm{fl}}$ will provide a new way of studying bosons
 critical behavior near the condensation transition, complementing recent
 studies of the critical correlation length
 \cite{Donner:2007,Dalibard:15}. Although $\chi_\mathrm{fl}$ can be detected
 for an ultracold gas in a usual harmonic trap \cite{Note4}, the most striking
 effect is expected for the gas in a uniform trap like the one recently
 implemented in Ref.~\cite{Hadzi,*hadzi2}, which is the most suitable platform
 for studying the BEC phase transition \cite{Dalibard:15}.  In this case we
 have found the unusually
sharp critical dependence of $\chi_{\mathrm{fl}}$ on $T-\Tc$,
 Eq.~(\ref{chi}) with   the corresponding critical exponent $\gamma$
equal   $2$ for $\tau\gtrsim\eta$.  Such a sharp criticality near the transition
 can provide another way to accurately determine \Tc. Finally, we hope that
 creating artificial magnetic fields with a constant gradient will
 find other applications in ultracold systems.

\begin{acknowledgements}
We are grateful to Peter Kr\"uger and Mike Gunn for useful discussions.
    One of us (I.V.L.) gratefully acknowledges support from  the Leverhulme Trust via the Grant No.\ RPG-380.
\end{acknowledgements}
%\bibliography{refs}\end{document}

\vspace*{18cm}
\newpage


\section*{Supplementary material}

# Supplemental Material

September 14, 2015

## 1 The chemical potential above $T_c$

We want to outline the reasoning that leads to the results in Eq. (6). Above the transition, nearly all the particles are in the excited states. This number is, at least above the transition, practically temperature independent as the number of particles in the ground state is vanishingly small. For a density of states of $\rho_\alpha(\epsilon) = C_\alpha \epsilon^{\alpha-1}$ one finds

$$n_{th} = \int d\epsilon \rho(\epsilon) \left(e^{\beta(\epsilon-\mu)} - 1\right)^{-1} = C_\alpha T^d \Gamma(\alpha) \, \mathrm{Li}_\alpha\left(e^{\frac{\mu}{T}}\right). \tag{1}$$

We used the definition of the polylogarithm

$$\mathrm{Li}_\alpha(\mathrm{x}) = \frac{1}{\Gamma(\alpha)} \int_0^\infty \frac{y^{\alpha-1}}{e^y/x - 1} dy. \tag{2}$$

At the critical temperature $T_c$, the chemical potential $\mu$ is 0. We have to demand that small changes in temperature $\Delta T = T_c \tau$ change $\mu$ in such a way as to leave the total particle number invariant. We use that $(T_c + \Delta T)^\alpha \approx T_c (1 + \alpha \tau)$ and the expansion of the polylogaritm[?]

$$\mathrm{Li}_\alpha(e^{\mathrm{x}}) = \Gamma(1-\alpha)(-\mathrm{x})^{\alpha-1} + \sum_k \frac{\zeta(\alpha-\mathrm{k})}{\mathrm{k}!} \mathrm{x}^{\mathrm{k}},$$

where $\zeta(r) = \sum_{n=1}^\infty \frac{1}{n^r}$ is the Riemann zeta function. This leads to

$$\tau = -\frac{1}{\alpha} \frac{\mathrm{Li}_\alpha\left(e^{\frac{\mu}{T}}\right) - \mathrm{Li}_\alpha(1)}{\mathrm{Li}_\alpha(1)}.$$

Since $\mu$ is small, we can focus on the relevant orders in the polynomial and obtain

$$-Li_\alpha(1)\alpha\tau = \mathrm{Li}_\alpha\left(e^{\frac{\mu}{T}}\right) = \Gamma(1-\alpha)\left(\frac{-\mu}{T_c}\right)^{\alpha-1} + \zeta(\alpha-1)\frac{\mu}{T_c}. \tag{3}$$

For the trapped gas $\alpha = 3$ and close enough to criticality the term quadratic in $\mu$ and logarithmic corrections can be neglected, one finds

$$\frac{\mu}{T_c} = -\frac{3\zeta(3)}{\zeta(2)}\tau = -C_1 T_c \tau \tag{4}$$

**Note:** It appears as if the polylogarithm diverges for integer $\alpha$ in the Gamma term. However, this divergence is exactly cancelled by the divergence of the $\zeta(1)$ term. A careful limiting procedure shows that for integer $k$ the limit $\lim_{\alpha \to k+1} \Gamma(1-\alpha)(-x)^{\alpha-1} + \frac{\zeta(\alpha-k)}{k!}x^k = \frac{x^k}{k!}\left(\sum_{n=1}^k \frac{1}{n} - \ln(-x)\right)$.

For $\alpha = 3/2$ (free uniform gas) and close to criticality, the linear in $\mu$ term can be neglected and one has

$$\frac{\mu}{T_c} = -\frac{1}{\pi}\left(\frac{3\zeta(3/2)}{4}\right)^2 \tau^2 = -C_2 T_c \tau^2. \tag{5}$$

## 2 The susceptibility in the uniform system

The following derivation is similar to the treatment of the subject in the context of superconductivity in [?]. Starting from the action in Eq. (5) we modify it to contain the magnetic vector potential $\mathbf{A} = \frac{\mathbf{r} \times \mathbf{B}}{2}$. In addition, the anisotropic stacked system (with spacing $d$) is introduced by replacing the gradient in the $z$ direction by $(\nabla_z \Psi)^2 \to J |\Psi_{n+1} - \Psi_n|^2$, where $n$ labels the layers. We work in the Gaussian approximation, so the quartic term proportional $g$ is neglected. To diagonalize, the Landau level representation is introduced

$$\Psi(\mathbf{r}) = \sum_{n,k_z} \Psi_{n,k_z} \phi_n(r_\parallel) e^{ik_z d},$$

where $r_\parallel$ is the position vector within one layer and $\phi_n$ is the wavefunction of the $n$th Landau level (which is degenerate with the factor $AB/\Phi_0$, $A$ being the sample area and $\Phi_0 = \hbar$ is the flux quantum) and $k_z$ is evaluated in the first Brillouin zone. Each state has the energy $T_c C_2 \tau^2 + \hbar \omega_B (n + 1/2) + 2J \cos(k_z d)$. The parameter $\omega_B = B/m$ is the cyclotron frequency, and $C_2$ is just the constants calculated in (5). As a reminder, the partition function $Z = \int D\Psi D\Psi^* e^{-S[\Psi, \Psi^*]}$ can be evaluated for a quadratic bosonic action by using the standard relation $\int D\Psi D\Psi^* e^{-\Psi^* A \Psi} = \det A^{-1}$. Taking the logarithm yields the formula for the free energy, $F = -T_c \ln Z$, given by

$$F(\tau, h) = \frac{AB}{\Phi_0} T_c \sum_{n,k_z} \ln \frac{\tau^{e_\alpha} + 4h(n + 1/2) + \frac{2J}{T_c C_\alpha}(1 - \cos(k_z l))}{\pi/c_\alpha}, \tag{6}$$

where $h$ is a reduced magnetic field $4h = \hbar B/mT_c C_2$. Since the sum is divergent one should introduce an upper cutoff $n_c \approx T_c/\hbar \omega_B \sim 1/h$ and for convenience it is useful to introduce the notation $\tilde{\tau} = \tau^2 + \frac{\eta}{2}(1 - \cos(k_z d))$, where $\eta = \frac{4J}{T_c C_2}$ is the anisotropy parameter. Exchanging the sum and logarithm and using [?]

$$\Gamma(z) = \lim_{n_c \to \infty} \frac{n_c! \, n_c^{z-1}}{z(z+1)(z+2) \cdots (z+n_c-1)}, \tag{7}$$

one obtains the approximation

$$F \sim \sum_{k_z} n_c \ln\left[\left(\frac{\pi}{4hc_\alpha}\right)\right] + \ln\left[\Gamma(1/2 + \tilde{\tau}/4h)\right] - \ln\left[n_c! \, n_c^{\tilde{\tau}/4h - 1/2}\right]. \tag{8}$$

Expanding to second order in $h$ and converting the sum over the momenta into an integral and keeping only the contribution $\sim h^2$ leaves

$$\delta F = \frac{ANT_c}{\pi \xi_0^2} \int_{-\pi/l}^{\pi/l} \frac{l dk_z}{2\pi} \frac{1}{3} \frac{h^2}{\tilde{\tau}} = \frac{ANTh^2}{3\pi \xi_0^2} \int_{-\pi}^{\pi} \frac{d\theta}{2\pi} \left(\tau^2 + \frac{\eta}{2}(1 - \cos(\theta))\right)^{-1}. \tag{9}$$

Performing the integration

$$\int_0^\pi \frac{d\theta}{2\pi} \left(\tau^2 + \frac{\eta}{2}(1 - \cos(\theta))\right)^{-1} = \left(\tau^2 (\tau^2 + \eta)\right)^{-1/2}. \tag{10}$$

With $V = ANl$ this gives for the free energy density

$$f(\tau) = \frac{F(\tau)}{V} = \frac{Th^2}{3\pi l \xi_0^2} \frac{1}{\sqrt{\tau^2 (\tau^2 + \eta)}} = -\frac{\chi}{2} B^2, \tag{11}$$

From which the susceptibility in Eq. (8) of the paper can be read off.

It is however helpful to rewrite everything in units of the energetic susceptibility per particle

$$\widetilde{\chi} = \frac{1}{NV} \frac{\partial^2 F}{\partial (\hbar \omega_B)^2} = \frac{m^2 \chi_{fl}}{\hbar^2} = \frac{1}{T_c} \frac{\sqrt{\eta}}{\sqrt{\tau^2 (\tau^2 + \eta)}}$$

where we used that $T_c \sim \frac{\hbar^2}{m l_{xy}^2} \sqrt{\eta}$ describes the suppression of the critical temperature due to the anisotropy, $l_{xy}$ being the interparticle distance within the layer. The above formula for the susceptibility interpolates between the strongly anisotropic($\approx$ single layer for $\tau \gg \eta$) and for the quasi three dimensional case ($\tau \ll \eta$). Note that the result differs from the similar result in superconductivity, namely here the divergence of the susceptibility is stronger. However, a system with an additional trapping in the $x-y$ direction behaves different. In the center of the trap where fluctuations are not influenced by the finite trap size, the correct susceptibility can be

recovered by letting $\tau^2 \to \tau$. However, this really holds only in the center of the trap and for not too small $\tau$. In the thermodynamic limit the fluctuational behaviour is suppressed, as we will demonstrate elsewhere.

One could speculate whether the Ginzburg criterion in the anisotropic system might considerably deviate from Eq. (7) of the paper. However, a similar rescaling as for the isotropic uniform system results in

$$Gi \approx 35 \left(\frac{mJd^2}{\hbar^2}\right)^{1/6} an^{1/3}.$$

Because of the small exponent, the Ginzburg criterion deviates only slightly for all values of practical interest. At this point it should be mentioned that in a real system the confinement may influence the effective scattering length.

## 3 Gauge potentials for multilevel systems

The following outline follows the review by **?** ].
In the conventional time-independent $\Lambda$ scheme, as exemplified by eq. (1) in the main text, the Hamiltonian $H$ can be written as

$$H = \left(\frac{P^2}{2m} + V(\mathbf{r})\right) \otimes \hat{1} + U,$$

where $V(\mathbf{r})$ is the trapping potential and $U$ is the spatially dependent matrix,

$$U = \frac{\hbar}{2}\begin{pmatrix} 0 & \Omega_1^* & 0 \\ \Omega_1 & 0 & \Omega_2 \\ 0 & \Omega_2^* & 0 \end{pmatrix},$$

that describes the coupling of the internal states $|g_{1/2}\rangle, |e\rangle$ via the light. The spatial dependence lies in the $\Omega_i(\mathbf{r})$. The matrix $U$ can be locally diagonalized by the states $|d(\mathbf{r})\rangle, |b^\pm(\mathbf{r})\rangle$, with

$$|d(\mathbf{r})\rangle = \frac{1}{\Omega}\left(\Omega_2 |g_1\rangle - \Omega_1 |g_2\rangle\right),$$

$$|b^\pm(\mathbf{r})\rangle = \frac{1}{\sqrt{2}}\left(\frac{\Omega_1^*}{\Omega}|g_1\rangle + \frac{\Omega_2^*}{\Omega}|g_2\rangle \pm |e\rangle\right).$$

The energy scale is given by $\Omega = \sqrt{|\Omega_1|^2 + |\Omega_2|^2}$ and the descriptions *dark state* for $|d\rangle$ and *bright states* for $|b^\pm\rangle$ are derived from the coupling properties of the states to the light and the excited state

$$U = \frac{\hbar\Omega}{2}\left(|b^+\rangle\langle b^+| - |b^-\rangle\langle b^-|\right).$$

The crucial point is that, after the preparation into one of the eigenstates, the atom moves only slowly in the light field. Then the internal state follows adiabatically the evolution of the state, say $|d\rangle$, and does not (appreciably) transit into any of the other two states, $|\pm\rangle$. The full state can at any point be represented in the spatially dependent basis $|d\rangle, |\pm\rangle$

$$|\Psi(\mathbf{r})\rangle = \sum_{\chi=d,\pm} \psi_\chi(\mathbf{r}) |\chi(\mathbf{r})\rangle. \tag{12}$$

Considering that

$$\nabla\left(\psi_\chi |\chi\rangle\right) = (\nabla\psi_\chi)|\chi\rangle + \psi_\chi |\nabla\chi\rangle, \tag{13}$$

one can determine the effective action of the momentum operator $P$. Projecting the Schrödinger equation onto the state $|d(\mathbf{r})\rangle$, one obtains the effective equation for the amplitude $\psi_d$

$$i\hbar\frac{\partial \psi_d}{\partial t} = \left[\frac{(\mathbf{P} - \mathbf{A}_{\text{art}})^2}{2m} + V + W_d\right]\psi_d.$$

The particle evolution of $\psi_d$ is described as the original equation of motion, but with an added vector potential $\mathbf{A}_{\text{art}} = i\hbar\langle d|\nabla d\rangle$ and the additional scalar potential $W = \hbar^2 |\langle b^+ + b^-|\nabla d\rangle|^2/2m$. In the following we assume that the trapping is stronger than the scalar potential; or, in the case of a uniform trap, that an external potential cancels its effect.

# 4 Detuning of the artificial magnetic field

Whereas the previous section outlined in very general terms how an artificial magnetic field is created, we want to describe here the more generalized setup from Fig. (1) in the main text of paper from the point of view of the dressed states of the atom. There are three internal states of the atom, two quasi-degenerate ground-states $|g_1\rangle, |g_2\rangle$ and one excited state $|e\rangle$, each having respectively the energy $\varepsilon_1, \varepsilon_2$ and $\varepsilon_e$ in the unperturbed Hamiltonian. The spatially dependent light field couples $|g_1\rangle$ with $|e\rangle$ with Rabi frequency $\Omega_1(\mathbf{r})e^{i\omega_1 t}$, $\omega_1$ being the frequency of the laser, whereas the second laser couples $|g_2\rangle$ with $|e\rangle$. The second laser is supposed to be a superposition of two light fields $\Omega_2 = \Omega_a(\mathbf{r})e^{i\omega_a t} + \Omega_b(\mathbf{r})e^{i\omega_b t}$, with respective frequencies $\omega_a$ and $\omega_b$. If we denote by $\omega_{e1} = (\varepsilon_e - \varepsilon_1)/\hbar$ and $\omega_{e2} = (\varepsilon_e - \varepsilon_2)/\hbar$ the Bohr frequencies between the levels, we obtain from the Schrödinger equation the set of equations for the $\tilde{b}_i$ of the state $|\psi\rangle = \tilde{b}_1 |g_1\rangle + \tilde{b}_2 |g_2\rangle + \tilde{b}_e |e\rangle$

$$i\hbar \frac{d}{dt}\begin{pmatrix} \tilde{b}_1 \\ \tilde{b}_e \\ \tilde{b}_2 \end{pmatrix} = \hbar \begin{pmatrix} \varepsilon_1/\hbar & \frac{\Omega_1}{2}e^{i\omega_1 t} & 0 \\ \frac{\Omega_1^*}{2}e^{-i\omega_1 t} & \varepsilon_e/\hbar & \frac{\Omega_a^*}{2}e^{-i\omega_a t} + \frac{\Omega_b^*}{2}e^{-i\omega_b t} \\ 0 & \frac{\Omega_a}{2}e^{i\omega_a t} + \frac{\Omega_b}{2}e^{i\omega_b t} & \varepsilon_2/\hbar \end{pmatrix} \begin{pmatrix} \tilde{b}_1 \\ \tilde{b}_e \\ \tilde{b}_2 \end{pmatrix} \quad (14)$$

We are looking first at the case when there are only two lasers coupling the three levels, one can always find a time-independent frame of reference. By introducing a slow moving variable $b_1$, one can separate out the fast moving component of the $\tilde{b}_i = e^{-i\delta_i t} b_i$, the $b_i$ can be made to be time independent. The $\delta_i$ have to be chosen appropriately. If the lasers are in resonance, $\omega_1 = \omega_{e1}$, $\omega_2 = \omega_{e2}$, then only $\delta_i = \epsilon_i/\hbar$ allows for true stationarity. The evolution of the $b_i$ can be described as the action of an effective Hamiltonian $H_{\text{eff}}$. In the resonant case it becomes

$$H_{\text{eff}} = \frac{\hbar}{2} \begin{pmatrix} 0 & \Omega_1^* & 0 \\ \Omega_1 & 0 & \Omega_2 \\ 0 & \Omega_2^* & 0 \end{pmatrix}., \quad (15)$$

which is equivalent to the formulation of eq. (1).

With $\Omega = \sqrt{|\Omega_1|^2 + |\Omega_2|^2}$ one finds the three eigenstates

$$|d\rangle = \frac{1}{\Omega}\left(-\Omega_2 |g_1\rangle + \Omega_1 |g_2\rangle\right), \qquad |b_\pm\rangle = \frac{1}{\sqrt{2}\Omega}\left(\Omega_1 |g_1\rangle \pm \Omega |e\rangle + \Omega_2 |g_2\rangle\right).$$

The dark state $|d\rangle$ has eigenenergy 0, whereas the two bright states $|b_\pm\rangle$ have energy $\pm\hbar\Omega/2$. This is just a reproduction of the situation in section 3.

Next we are trying to find the asymptotics for the artificial magnetic field for small and large values of $\delta/\Omega$.

If not all the laser are tuned to resonance, there is a certain freedom in choosing the individual $\delta_i$ to achieve this. The physics are however not affected by such a choice.

For convenience we choose it to act as a perturbation to the Hamiltonian of the form

$$\delta V = \delta \begin{pmatrix} -1 & 0 & 0 \\ 0 & 0 & 0 \\ 0 & 0 & +1 \end{pmatrix}.$$

Here $\delta$ is the small (with respect to $\Omega$) detuning. Using standard perturbation theory, the dark state is changed in leading order (for $|\Omega_1| = |\Omega_2| = \Omega/\sqrt{2}$) to

$$|d\rangle_\delta = \left(1 + \frac{\delta^2}{2\Omega^2}\right)^{-1/2} \left(|d\rangle + \frac{\delta\left(\Omega_1\Omega_2^* + \Omega_1^*\Omega_2\right)}{\sqrt{2}\Omega^3} \left(|b_+\rangle - |b_-\rangle\right)\right).$$

The resulting magnetic potential for a system prepared in the dark state that movies adiabatically in the Rabi fields is given as (see [?]) $\mathbf{A}_{\text{art}} = i\langle d|\nabla d\rangle$. The light fields are supposed to be of the Gauss-Laguerre type $\Omega_i = \frac{\Omega}{\sqrt{2}}\left(\frac{\rho}{\rho_0}\right)^{l_i} e^{il_i\phi}$, where $\rho$ is the radial coordinate and $\phi$ the angular one. Especially, since we are interested in the magnetic field in the $z$ direction we are looking for $A_{\text{art},\phi} = \langle d|\frac{1}{\rho}\partial_\phi|d\rangle$

It becomes clear, from the form of the bright states, that $\partial_\phi |b_+\rangle = \partial_\phi |b_-\rangle$, as the term in which the two differ, $\pm\Omega |e\rangle$, is $\phi$ independent. Because both, $|b_\pm\rangle$ perturb $|d\rangle$ with an opposite sign, their direct contribution to the vector potential of the ground state cancels and the only change to the vector field comes from the lower weight of the dark state, aka $\frac{1}{\rho}\partial_\rho |d\rangle_\delta = \left(1 + \frac{\delta^2}{2\Omega^2}\right)^{-1/2} \frac{1}{\rho}\partial_\rho |d\rangle$. Hence $B_{\text{art}}(\delta) \approx \frac{2\Omega^2}{2\Omega^2+\delta^2} B_{0,\text{art}}$, where $B_{0,\text{art}}$ is the magnetic field of the dark state on resonance.

In the opposite limit and for large detuning ($\delta/\Omega \gg 1$) the unperturbed Hamiltonian is

$$H_0 = \hbar \begin{pmatrix} -\delta & 0 & 0 \\ 0 & \delta_e & 0 \\ 0 & 0 & \delta \end{pmatrix},$$

where the eigenstates are just the original states $|g_1\rangle, |g_2\rangle |e\rangle$. Adding now a small coupling via the Rabi fields, one obtains perturbations of the form

$$|e\rangle_\delta = \left(1 + \frac{2\Omega^2}{(\delta_e + \delta)^2}\right)^{-1/2} \left(|e\rangle + \frac{\Omega_1}{\delta - \delta_e} |g_1\rangle + \frac{\Omega_2}{\delta_e - \delta} |g_2\rangle\right).$$

Similar states appear for the perturbation of the other two states. The influence on the magnetic field can be easiest seen in the case where only $\Omega_2$ is angular dependent. Then the strength of the magnetic vector potential $\sim \langle e|\frac{1}{\rho}\partial_\phi e\rangle$ will be dominated by the prefactor of $|g_2\rangle$ state and the magnetic field will be of the order $\Omega^2/\delta^2$ or $B_{\text{art}}(\mathbf{r}, \delta) \sim \frac{2\Omega^2}{\delta^2} B_{0,\text{art}}(\mathbf{r})$, just as a naive extension of the weak detuning result would suggest.

These findings, namely that $B_{\text{art}}(\delta) \approx \frac{2\Omega^2}{2\Omega^2 + \delta^2} B_{0,\text{art}}$ for $\delta/\Omega \ll 1$; and $B_{\text{art}}(\mathbf{r}, \delta) \sim \frac{2\Omega^2}{\delta^2} B_{0,\text{art}}(\mathbf{r})$ for $\delta/\Omega \gg 1$, suggest that the magnetic field is an analytic function of $\delta/\Omega$ and that the magnetic field can be written as

$$B_{\text{art}}(\mathbf{r}, \delta) = B_{0,\text{art}}(\mathbf{r}) f(\delta/\Omega),$$

as in eq. (4) of the paper.

Now we are looking at the case when the third laser is turned on $\Omega_b \neq 0$, but is also far detuned from the other laser on the same transition $\delta_a - \delta_b \gg \Omega$. The Hamiltonian becomes time-dependent when $\Omega_b \neq 0, \omega_b \neq \omega_a$. The time dependance cannot be eliminated by a rotation of the states, one will always have a system that oscillates with frequency $\Delta = \delta_a - \delta_b$. One can generalize the previous decomposition of the solution into $b_1, b_2$ and $b_e$. Apart from an overall phase evolution, each of the $c_i$ should be a periodic function of time, with period given by $T = \frac{2\pi}{\Delta}$ (Floquet theorem or Bloch theorem for a time dimension). Formally one can Fourier expand the $c_i(t)$ to contain all the overtones, e.g. $b_1 |g_1\rangle \to \sum_n b_1^{(n)} e^{i\Delta n t} |g_1, n\rangle$. Quite generally, the state of the atom can be decomposed into a Floquet basis, $|\psi\rangle = e^{i\varepsilon t} \sum_{n,i} c_i^{(n)} e^{i\Delta n t} |i\rangle$, where $n$ is an integer of the Floquet basis and $i$ is the index of the standard set of $|g_1\rangle, |g_2\rangle$ and $|e\rangle$. For very large detuning, only the time-independent $n = 0$ subspace is relevant. However by slowly decreasing the detuning, the other Floquet sectors come into play. The stationary Schrödinger equation for the slow variables $b_i^n$ is (given $\Omega_a$ in in resonance)

$$(\varepsilon + n\Delta) b_1^{(n)} = \Omega_1 b_e^{(n)}$$
$$(\varepsilon + n\Delta) b_2^{(n)} = \Omega_a b_e^{(n)} + \Omega_b b_e^{(n-1)}$$
$$(\varepsilon + n\Delta) b_e^{(n)} = \Omega_1^* b_1^{(n)} + \Omega_a^* b_2^{(n)} + \Omega_b^* b_2^{(n+1)}.$$

The perturbation $V = \sum_n \Omega_b |g_2, n\rangle \langle e, n-1| + \Omega_3^* |e, n\rangle \langle g_2, n+1|$ couples the different blocks indexed by $n$, each with (large) offset $n\Delta$. Without the coupling, every block can be diagonalized into the same set of eigenstates just as the time-independent problem without detuning. The coupling between different blocks is

$$V_{n,n+1} = \sum_{i,j} |i, n\rangle \langle i, n| V |n+1, j\rangle \langle n+1, j| = \Omega_b^* |e, n\rangle \langle g_2, n+1|,$$
$$V_{n+1,n} = \Omega_b |g_2, n+1\rangle \langle e, n|.$$

It is notationally convenient to separate the two indices i.e. $|i, n\rangle \to |i\rangle |n\rangle$, as we are interested in the coupling between the $n$. If one starts with a single laser field $\Omega_a$ in resonance and in a dark state ($|d\rangle |n = 0\rangle$) and introduces the second light field by adiabatically lowering the detuning $\Delta = \infty$ to a finite (but large) $\Delta$, one obtains to lowest order in perturbation theory

$$|\phi^1\rangle \approx \begin{pmatrix} -\frac{\Omega_2^*}{\Omega} \\ 0 \\ \frac{\Omega_1^*}{\Omega} \end{pmatrix} |n = 0\rangle - \frac{\Omega_b^*}{\Delta^2} \begin{pmatrix} \frac{|\Omega_1|^2}{\Omega} \\ \frac{\Delta\Omega_1^*}{\Omega} \\ \frac{\Omega_1^*\Omega_a}{\Omega} \end{pmatrix} |n = -1\rangle.$$

where we used that $\frac{1}{\Delta^2 - \Omega_2^2} \approx \frac{1}{\Delta^2}$. The state obtains a small component that is oscillating but also carries the spatial information necessary for an artificial magnetic field. The magnetic vector potential to leading order in $\frac{\Omega_b^*}{\Delta}$ can be obtained as before by calculating the overlap $\langle \phi^1 | \nabla \phi^1 \rangle$ (after normalization). The perturbation adds an additional term to the magnetic vector potential proportional to the angular momentum of the beam $\Omega_b e^{il_b\phi}$

5

and it is the same contribution as in the previous case of a single far-detuned laser, since $\Omega_3 \approx \Omega$. Therefore this amounts to effectively adding the magnetic fields created by the two Rabi fields individually.

One can argue that this perturbative expansion breaks down for $\Omega/\Delta \approx 1$. However, by choosing the angular momenta of the beams sufficiently different, one can make the value of $\Omega/\Delta$, for which the second derivative of the magnetic field vanishes and the gradient is approximately constant over an interval of order $\Omega$ as small as wanted, thereby making sure that the perturbation theory stays valid at all times. If one of the fields is in resonance, then the detuning has two effects. On the one hand the magnetic field created by the first field weakens. On the other hand, the atom gets closer to resonance with the second field, which is stronger by the factor $\left(\frac{l_b}{l_a}\right)^2$. In particular, the difference between $\delta_a$ and $\delta_b$ can be chosen such that downward curvature from the detuning with the first laser is compensated by the opposite curvature of the second field, thereby creating a steady gradient.

Simple plotting shows, that for $\Delta l = 1$, the plateau of the gradient forms at $\Omega/\Delta = 0.7$, for $\Delta l = 2$ at $\Omega/\Delta = 0.4$ and for $\Delta l = 3$ at $\Delta/\Omega = 0.32$.

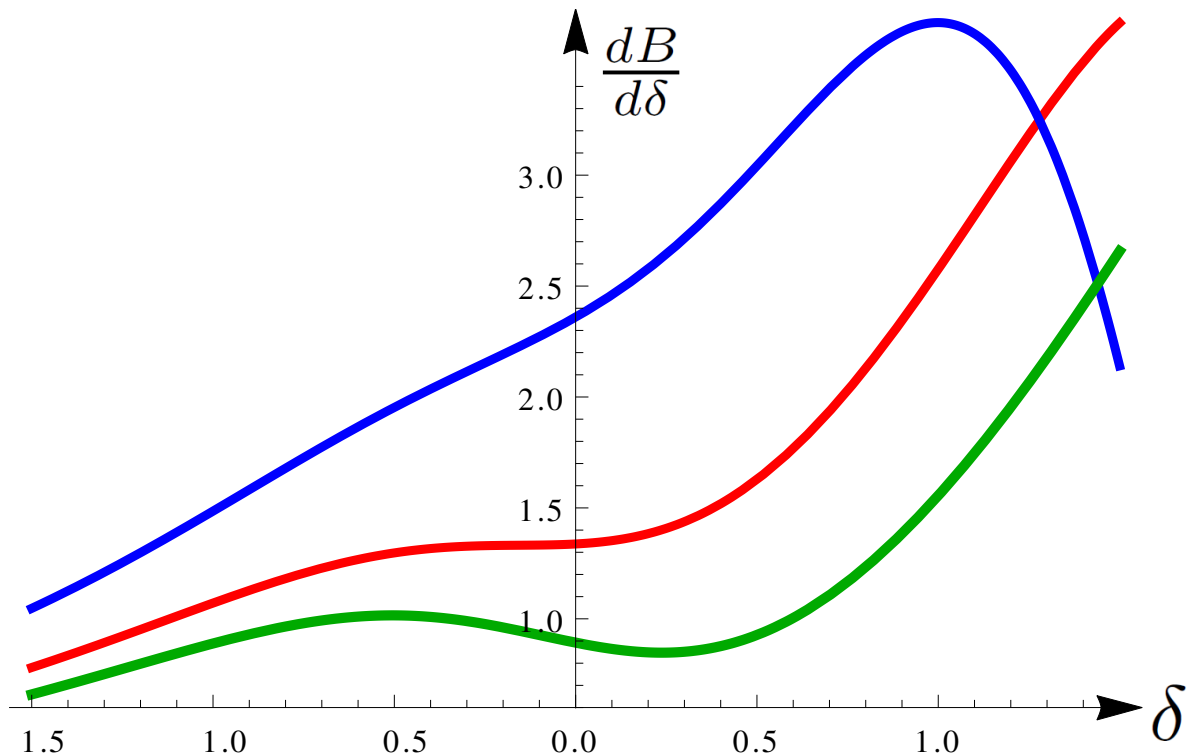

Figure 1: The gradient of the magnetic field $\frac{dB}{d\delta}$ for different detunings $\Delta/\Omega$ of two beams with $\Delta l = 2$. In the upper graph (blue) $\Delta/\Omega \approx 2.0$. In the middle graph (red) $\Delta/\Omega \approx 2.5$ and in the lower graph (green) $\Delta/\Omega \approx 2.8$. One can clearly see how the finetuning of $\Delta/\Omega$ allows for the compensation of curvatures by creating a plateau of width $\sim \Omega$.



\end{document}